# Confining deep eutectic solvents in nanopores: insight into thermodynamics and chemical activity

Benjamin Malfait, Aicha Jani, Denis Morineau [*]

[†]Institute of Physics of Rennes, CNRS-University of Rennes 1, UMR 6251, F-35042 Rennes, France

ABSTRACT

We have established the detailed phase diagram of the prototypical deep eutectic solvent ethaline (ethylene glycol / choline chloride 2:1) as a function of the hydration level, in the bulk state and confined in the nanochannels of mesostructured porous silica matrices MCM-41 and SBA-15, with pore radii $R_P$ = 1.8 nm and 4.15 nm. For neat and moderately hydrated DESs, freezing was avoided and glassforming solutions were formed in all cases. For mass fraction of water above a threshold value $W_g$' $\approx$ 30%, crystallization occurred and led to the formation of a maximally-freeze-concentrated DES solution. In this case, extremely deep melting depressions were attained in the confined states, due to the combination of confinement and cryoscopic effects. These phenomena were analyzed quantitatively, based on an extended version of the classical Gibbs-Thomson and Raoult thermodynamic approaches. In this framework, the predicted values of the water chemical activity in the confined systems were shown to systematically deviate from those of the bulk counterparts. The origin of this striking observation is discussed with respect to thermodynamic anomalies of water in the 'no-man's land' and to the probable existence of specific nanostructures in DES solutions when manipulated in nanochannels or at interfaces with solids.

**Corresponding Author :** * E-mail: **denis.morineau@univ-rennes1.fr**

**KEYWORDS** Deep eutectic solvents, Ethaline, Confinement, Mesoporous silica, MCM-41, SBA-15, Water chemical activity, Differential scanning calorimetry

# 1. Introduction

For the last decade, deep eutectic solvents (DESs) have emerged as promising alternative to classical solvents and have received increased attention for practical applications in various industrial fields.[1-6] DESs are often obtained by mixing of an H-bond donor (HBD) with a molecular ion or H-bond acceptor (HBA).[6] A significant lowering of the melting point is obtained for compositions around the eutectic point, which makes it possible to design countless sorts of solvents, liquid at room temperature, by combining different components which would otherwise be solid.[7, 8] Non-ideal mixing effects related to H-bonds, ionic interactions, and supermolecular correlations between the DES constituents are also often associated to exceptionally deep depression of the melting point.[9-11] An upper level of complexity is attained when DESs are used as ingredients of multicomponent systems. The most significant case consists in mixing DES and water, which advantageously lowers the viscosity of the solvent and makes it also possible to adjust many other physical properties, such as the conductivity, the dielectric permittivity, the dipolar relaxation dynamics and the mesoscopic structure, while possibly retaining its functional properties. [11-23]

For many applications related to catalysis, electrochemistry, gas capture, and energy storage, solvents are not considered as bulk liquid phases but handled at solid interfaces or in mesoporous hosts.[24-26] However, when they are spatially confined at the nanometer scale, several fundamental properties of liquids considerably differ from those of their bulk counterparts. While confinement effects have been investigated for classical molecular liquids in numerous studies, [27-33] the topics has been virtually unexplored for DESs. [34-36]

In the present work, the confinement effects on the phase behavior of aqueous mixtures of the prototypical DES named ethaline (ethylene glycol/choline chloride) have been established experimentally by differential scanning calorimetry (DSC) for an extended range of compositions, from mild hydration levels to complete dilution, and for temperatures ranging from 120 to 290 K. The selection of ethaline-water system was prompted by recent studies that provide an in-depth viewpoint on its entire phase diagram, [22] ionic conductivity and dipolar reorientational dynamics by broadband dielectric spectroscopy, [23] and its mesoscopic structural organization by neutron scattering [11] and molecular simulation.[10] The experimental realization of spatial confinement was achieved thanks to the well-defined cylindrical pores of mesostructured silicas MCM-41 and SBA-15 (pore radius $R_P$ = 1.8 and 4.15 nm). As a whole, this work addresses for the first time the cumulative effects of confinement (Gibbs-Thomson effect) and solutes concentration (Raoult's law) on the freezing depression of confined DES aqueous solutions, and it aims at extending the latest developments made in the field for glycerol aqueous solutions.[37]

## 2. Samples and methods

Choline chloride (>99%) and ethylene glycol (anhydrous, 99.8%) were purchased from Sigma-Aldrich. Ethaline DES was prepared by weighting and adding ethylene glycol and choline chloride in a molar ratio of 2:1, mixed by mechanical agitation at about 60°C for 30 min until a clear homogeneous liquid phase was obtained and served as stock solutions. Series of 11 working solutions was prepared by pipetting and addition of ultrapure water (18.2 MΩ.cm) for regularly spaced mass fractions $W$=0-100%.

The mesoporous materials MCM-41 silicas were prepared in our laboratory according to a procedure similar to that described elsewhere [38] and already used in previous works.[28, 39-41]

Hexadecyl-ammonium bromide was used as template to get a mesostructured hexagonal array of aligned channels with pore radius $R_P = 1.8$ nm as determined by nitrogen physisorption. The SBA-15 mesoporous silicas were prepared using a procedure described elsewhere,[39, 42-45] with slight modifications of the thermal treatments to optimize the final structure of the product.[46] Nonionic triblock copolymer (Pluronic $P_{123}$): $(EO)_{20}(PO)_{70}(EO)_{20}$ was used as a template to obtain a mesostructured hexagonal array of aligned channels with a pore radius $R_P = 4.15$ nm.

The calcined matrices were dried at 120°C under primary vacuum for 12 hours prior to the experiments. The empty MCM-41 and SBA-15 (from 8 to 10 mg) were packed in DSC Tzero© aluminum hermetic pans and then filled by liquid imbibition with DES aqueous solution injected from a syringe. The weighted amount of DES aqueous solution was determined precisely to allow loading of the porous volume with no excess liquid out of the porosity. It corresponded to a filling ratio $\rho$ (expressed as liquid volume per mass of matrix) of $\rho = 0.63 \pm 0.02$ $cm^3.g^{-1}$ and $\rho = 0.83 \pm 0.02$ $cm^3.g^{-1}$ for MCM-41 and SBA-15, respectively. We verified for a limited number of compositions that, for slightly overfilled samples, the presence of excess liquid could be detected by additional thermal events located at temperatures corresponding to bulk transitions, which is also in accordance with previous DSC experiments.[37, 47]

The differential scanning calorimetry (DSC) measurements were performed with a Q-20 TA Instrument equipped with a liquid nitrogen cooling system. The melting transition of an indium sample was used for calibration of temperature and heat flux. In general, the temperature was ramped linearly on cooling and heating in the temperature range from 120 to 290 K (scanning rate of 10 $K.min^{-1}$). For samples with water mass fraction $W = 40\%$ and 50%, additional thermal pretreatments were also investigated such as quenching down to 120 K (in order to limit crystallization) or cycling in the range 120-170 K (in order to favor complete crystallization), and thereafter followed by a final heating branch from 120 to 290 K at 10 K.min$^{-1}$.

## 3. Results and discussion

*3.1 Thermograms*

The thermograms measured on heating are shown in a 3D representation in Fig. 1 for bulk and confined DES aqueous solutions. Two distinct thermal events are visible : a jump in the heat flux, at temperature about 150 K, which is the signature of a glass transition, and an endothermic peak, located in the temperature range from 230 to 280 K, that is related to melting.

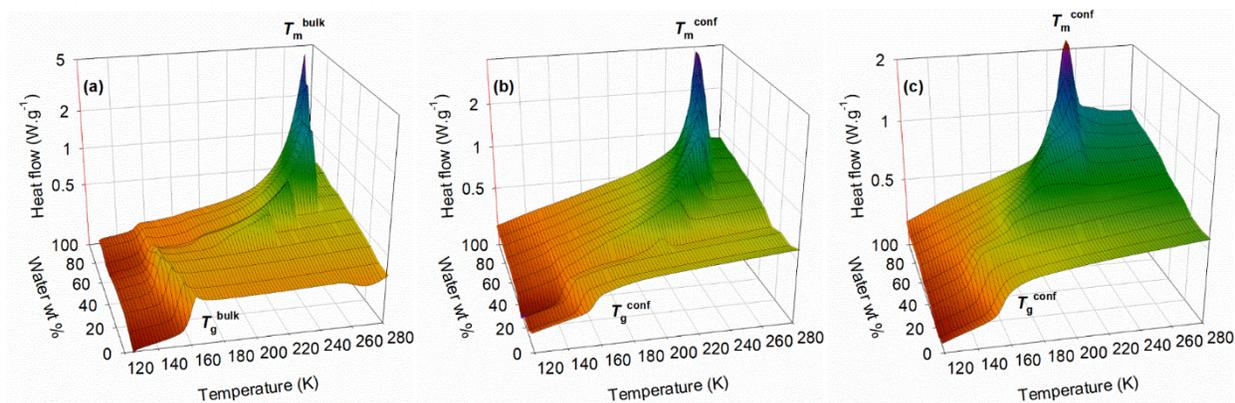

**Figure 1**. 3D representation of the thermograms measured during heating for of aqueous solutions of Ethaline (2:1) DES mixtures as a function of the water mass fraction. The heat flow (endotherm up) is represented in log-scale for better clarity. The glass transition is indicated by a jump of the heat capacity at $T_g$ and the melting transition by an endothermic peak at $T_m$. (a) bulk DES solutions adapted from [22], (b) DES solutions confined in SBA-15 with pore size $R_P$ = 4.15 nm, and (c) DES solutions confined in MCM-41 with pore size $R_P$ = 1.8 nm.

A detailed discussion of the phase behavior has been performed for bulk aqueous solutions of ethaline by Jani *et al.*.[22] Two different regimes were observed depending on the hydration level, with respect to a threshold value $W_g' \approx 30\%$. For low hydration levels ($W \leq W_g'$), no crystallization/melting was observed on cooling/heating under the experimental conditions. In fact in this case, corresponding to 'Water-in-DES' according to ref. [19], the entire solution vitrifies. This is confirmed by the single thermal signature of the glass transition. The observations made for DES solutions in the two confined geometries were qualitatively the same as those reported for bulk solvents. The minor exception concerns the solution with $W = 30\%$ confined in SBA-15, that presented a weak endothermic peak (visible in Fig. 1b), indicating that a small fraction of ice formed. This observation is likely due to the close proximity of $W$ with the threshold value $W_g'$.

For $W > W_g'$, both crystallization/melting and glass transition phenomena were observed. It indicates that in this case, denoted 'DES-in-Water', ice coexists with a freeze-concentrated aqueous solution of DES at low temperature. In the bulk, for $W = 40$-$50\%$, it was however also possible to avoid crystallization during fast cooling and maintain the entire solution in a homogeneous supercooled metastable or vitreous non-equilibrium state. In confinement, for $W = 40$-$50\%$, crystallization always occurred on cooling at 10 K.min$^{-1}$. However, on re-heating, we also observed a tiny residual exothermic peak (representing $\approx 3\%$ of the total crystallization enthalpy) at about $T_c = 170$ K (cf. Fig. S1). We attribute this feature to cold crystallization of a small fraction of metastable freezable water that remained trapped in the liquid solution during the cooling ramp. This phenomenon indicates slow crystallization kinetics, and as such, it could indeed be circumvented by applying a thermal cycle in the range 120-170 K before final heating from 120 to 290 K (cf. Fig. S1).

Despite the apparent qualitative similarity between the phase behavior of bulk and confined DES solutions, remarkable differences in transition temperatures and their dependence on hydration level are visible in Fig. 1. In order to better exemplify the respective effects of confinement and hydration on the phase transitions, a selection of thermograms are compared in Fig. 2.

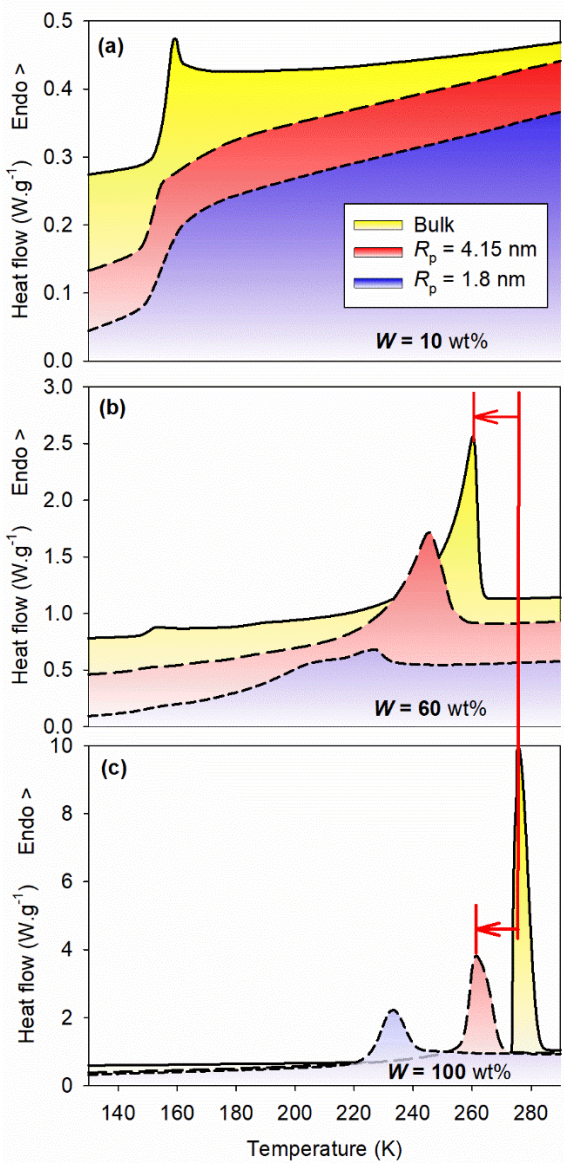

**Figure 2**. Confinement effect on the thermograms of aqueous solutions of Ethaline (2:1) DES mixtures measured in the bulk (solid line with yellow-filled area), in SBA-15 (long dashed line with red-filled area), and in MCM-41 (short dashed line with blue-filled area). Water mass fractions are (a) $W = 10\%$, (b) $W = 60\%$, and (c) $W = 100\%$ (pure water). In panel (c), the red arrow illustrates the melting peak depression due to solely the Gibbs-Thomson effect, while in panel (b), it illustrates the melting peak depression due to solely the cryoscopic (Raoult's law) effect.

Fig. 2a illustrates the confinement effect on the glass transition of the 'Water-in-DES' solution, corresponding to a hydration level, $W = 10\%$, smaller than $W_g'$. For bulk ethaline solution, the heat capacity jump, around the midpoint $T_g = 155$ K, is rather sharp (i.e. occurring on a few degrees). It is also associated with an endothermic overshoot that is a characteristic feature of glass-forming liquids, and that we attribute to enthalpy relaxation in the vicinity of the glass transition. [48] In confinement, the glass transition is spread out, with a temperature broadening that is larger the smaller the pore size. For prototypical glassforming liquids, such as toluene, similar confinement effects were attributed to the broadening of the distribution of relaxation times due to surface-induced spatial dynamic heterogeneities.[28] It is also worth noting that, although occurring over a wide temperature range, the glass transition of confined DES aqueous solutions occurred in a single step. It means that expected supermolecular entities formed by strong attractive interactions in DESs do not lead to the observation of distinct calorimetric glassy dynamics.[7, 9-12, 16] This is different from recent work on organic binary solvents which, when confined, present microphase-separated structures with two distinct molecular dynamics and distinct glass transitions. [49-52] Concerning the average glass transition temperature, it was found weakly affected by confinement, but with a larger downward shift for SBA-15. For $W = 10\%$, it was $T_g = 151.5$ K for DES in SBA-15, compared to respectively 155.8 K and 155.4 K, for DES in MCM-41 and in the bulk state. This means that the dependence of $T_g$ on pore size $R_P$ is non-monotonic. Such a variation has been reported for different glassforming systems, and attributed to competing interfacial and finite size effects on glassy dynamics.[28, 53]

We now discuss the melting transition for the 'DES-in-water' regime ($W > W_g'$), which is illustrated in Fig. 2b for $W = 60\%$ and compared to pure water in Fig. 2c. As a general trend, it is obvious that the melting point decreased when $W$ and/or $R_P$ decreased. The first phenomenon, classically denoted cryoscopic effect, expresses the effect of adding a solute on the freezing/melting behavior of the solvent. It was rationalized in the late 19th century by one of the three Raoult's laws, which embrace the overall colligative properties of solutions.[54] For bulk DES solutions, the cryoscopic depression of the melting point for $W = 60\%$ is 20 K, as indicated by an arrow in Fig. 2b. The second phenomenon, denoted confinement effect, is classically denoted as the Gibbs-Thomson effect.[55] It expresses the growing importance of interfacial energy in determining the thermodynamic stability of crystal with nanometer scale dimension. For pure water, the Gibbs-Thomson effect due to confinement with $R_P = 4.15$ nm is indicated by an arrow in Fig. 2c. Interestingly, the selected systems (composition and pore size) highlight that it may easily happen that the melting depressions due to both phenomena, when considered separately, (i.e. cryoscopic and Gibbs-Thomson effects) have about the same magnitude.

*3.2 Phase diagram*

When addressing the interesting case of confined DESs aqueous solutions, the question emerges about how these two effects actually combine. To address this issue, we present in Fig. 3 the phase diagram of bulk and confined solutions. It should be stressed however, that the studied systems are actually ternary mixtures, and only a pseudo-binary phase diagram along the line of constant ethylene glycol/choline chloride relative mole ratio was determined.

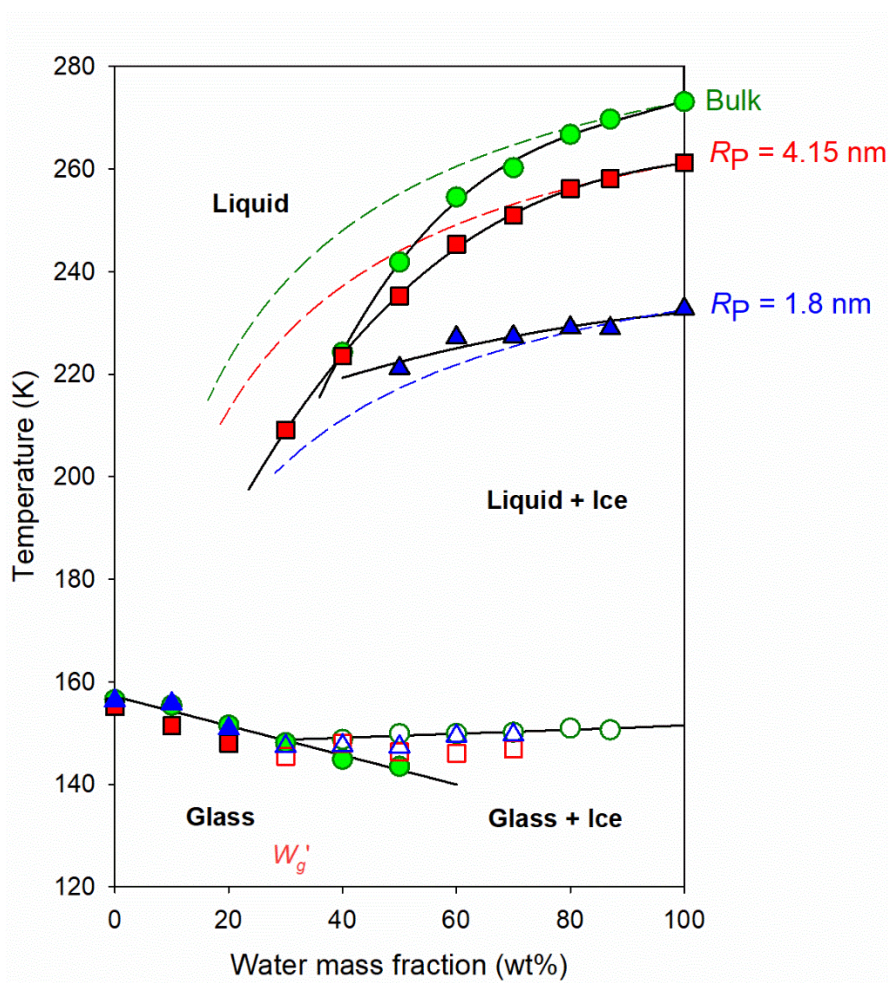

**Figure 3**. Phase diagram of aqueous solutions of ethaline DES measured in the bulk state (green circles), confined in SBA-15 (red squares), and confined in MCM-41 (blue triangles). In the high

temperature part (above 200 K) the liquidus lines are located in water rich-region ($W > 30\%$). In the low temperature part (below 160 K), the glass transition line of the entire solution (filled symbols), and the glass transition line of the maximally-freeze concentrated solution co-existing with ice (open symbols) intercept at $W_g' \approx 30\%$. Dashed lines are theoretical predictions based on classical thermodynamics (see text).

For $W < 30\%$, the glass transition of the aqueous DES solution decreases steadily with increasing the hydration level. In the bulk, this tendency can be correlated to the decrease of the viscosity and increased molecular mobility.[23] Interestingly, we can conclude that this beneficial effect of water on the dynamics of DES is retained in nanoscopic environment. For $W = 30\text{-}90\%$, the glass transition temperature assumed an almost fixed value. This indicates that, after water crystallization, the composition of the unfrozen glassforming solution is rather constant and independent on the overall composition $W$. This interpretation found an independent confirmation provided by the melting enthalpy of $\Delta H_m$, which can be considered as a measure of the fraction of ice formed.[22] As illustrated in Fig. 4, the values of $\Delta H_m$ follow a linear dependence on $W$ and they can be extrapolated at zero enthalpy to a non-vanishing value of $W$. We can therefore conclude that, in the 'DES-in-Water' regime, crystallization leads to the formation of a maximally-freeze concentrated DES solution with a composition $W_g' = 30 \pm 5\%$ that is comparable in the bulk and confined states.

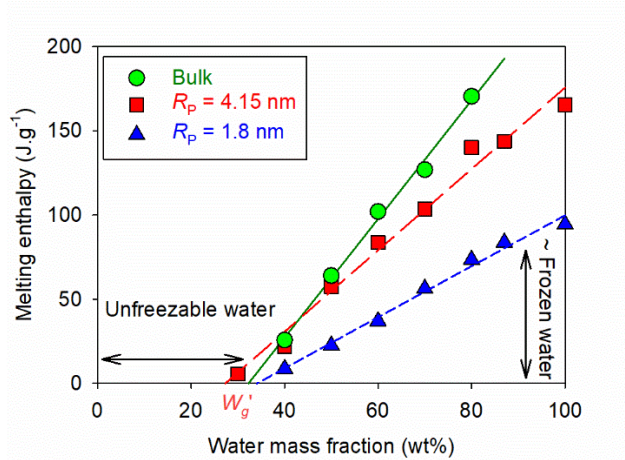

**Figure 4**. Variation of the melting peak enthalpy with the water mass fraction for aqueous solutions of ethaline DES measured in the bulk state (green circles), confined in SBA-15 (red squares), and confined in MCM-41 (blue triangles). The linear extrapolations at zero enthalpy converge towards the value of the maximally-freeze concentrated solution, $W_g' \approx 30\%$.

*3.3 Thermodynamic modeling of liquidus and chemical activity*

During heating, for $W$ = 30-90%, the melting process occurs on a broad temperature range, which ends at the liquidus line. In confined geometry, the liquidus temperature classically corresponds to the melting point of a crystal of ice of nanometer size $r$ in equilibrium with a surrounding solution of composition $W$. [37] It has been shown for confined pure water that $r$ is closely related to the radius of the pore, such that $r = R_p - e$, if one takes into account an unfrozen interfacial layer of thickness $e \approx 0.6$ nm. [55] In this situation, it can be demonstrated from classical thermodynamics that the melting point $T_r^W$ is determined by eq. 1, where $\gamma_{sl}$ and $v_s$ are the ice-liquid surface tension and ice molar volume, $a$ is the water chemical activity in the DES aqueous solution and $R$ the gas constant.[37] The quantity $\Delta\mu_0(T) = \mu_s(T) - \mu_l(T)$ is the difference between the chemical potential of bulk ice and bulk pure liquid water at temperature $T$.

$$\Delta\mu_0(T_r^W) = -2\frac{v_s}{r}\gamma_{sl} + RT_r^W \ln(a) \qquad (1)$$

For temperature close to the bulk melting point of pure water, $T_{bulk}^0 = 273{,}15$ K, $\Delta\mu_0(T)$ can be approximated by

$$\Delta\mu_0(T) \approx \widehat{\Delta\mu_0}(T) = -\Delta H_m\left(1 - \frac{T}{T_{bulk}^0}\right) \qquad (2)$$

with $\Delta H_m$ being the melting enthalpy of ice (cf. Supporting Information). This corresponds to the level of approximation classically made to derive the Gibbs-Thomson and the Raoult's laws. For confined aqueous solutions, the liquidus temperature $T_r^W$ writes then as an extended version of the Gibbs-Thomson equation

$$T_r^W - T_{bulk}^0 = \frac{-2\gamma_{sl}v_s T_{bulk}^0}{\Delta H_m r} + \frac{RT_r^W T_{bulk}^0 \ln(a)}{\Delta H_m} \qquad (3)$$

Applied to pure water, this equation was found in good agreement with the experimentally determined melting points (cf. Tables S1 and S2). If one assumes that $\gamma_{sl}$ is not significantly dependent on the liquid composition, the variation of the liquidus temperature $T_r^W$ with the concentration of the aqueous DES solution can be determined by

$$\Delta\mu_0(T_r^W) = \Delta\mu_0(T_r^0) + RT_r^W \ln(a) \qquad (4)$$

which is simplified to the following extended cryoscopic equation for confined systems when the approximated version of $\widehat{\Delta\mu_0}(T)$ is used.

$$\Delta H_m \left(\frac{T_r^w - T_r^0}{T_{bulk}^0}\right) = RT_r^w \ln(a) \qquad (5)$$

The predicted liquidus lines obtained from Eq. 5 in the limit of ideal mixing approximation are indicated in Fig. 3 as dashed lines. They are in fair agreement with experimentally determined melting points for large hydration levels, but significant deviations are observed for smaller hydration, typically for $W < 70\%$. The experimental melting point depression is larger than predicted for aqueous DES solutions in the bulk state and in SBA-15. An opposite situation is obtained for stronger confining condition in MCM-41. We have identified two possible origins of the observed discrepancy that are not specific of the confined systems. Firstly, the classical first-order approximation of the chemical potential of bulk water around $T_{bulk}^0$ (Eq. 2) neglects the contribution from heat capacities and their temperature dependences. Its validity is certainly questionable for the DES systems studied in the present work that demonstrate exceptionally deep melting depression, up to - 50 K. Secondly, deviations from ideality have been commonly reported for DES and they certainly need to be accounted for in the case of DES aqueous solutions.[7]

In order to incorporate these two elements in the thermodynamic description, we have evaluated the water chemical activity from eq. 4, based on the experimentally determined liquidus temperatures and using for $\Delta\mu_0(T)$ an improved description of the water Gibbs free energy, determined by Johari *et al*. [56] and later parametrized in a temperature range from 273,15 K to 153 K by Koop *et al*. [57] (cf. eq. S3). We present the resulting estimates of the water chemical activity as a function of the composition in Fig. 5. The water activity presents a linear dependence on the water molar fraction, with slope that depends on the conditions of confinement. For bulk DES aqueous solutions, the activity coefficient is smaller than one. This most probably reflects specific interactions between water and the DES components, as already discussed from the structural point of view.[10-12] Interestingly, the as-obtained values of water activity are consistent with very recent determinations made for a restricted range of composition, using a different approach based on the partial pressure of the vapor phase as illustrated by open diamonds in Fig. 5b [58].

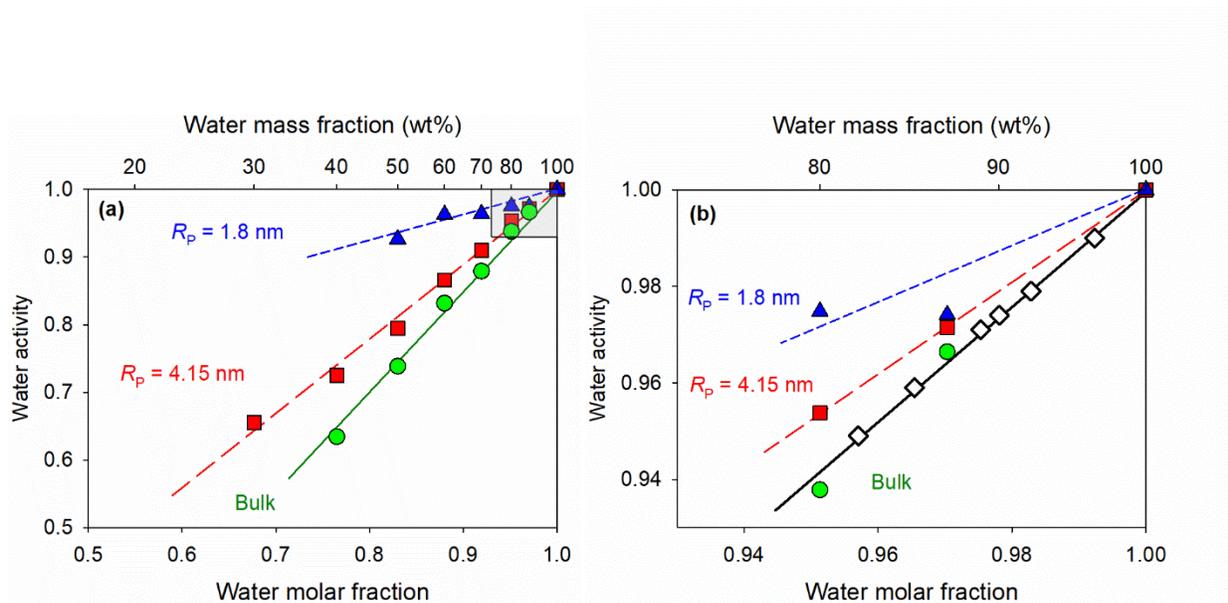

**Figure 5**. Water chemical activity in aqueous solutions of ethaline DES in the bulk state (green circles), confined in SBA-15 (red squares), and confined in MCM-41 (blue triangles) obtained from the experimental melting points and using an extended cryoscopic thermodynamic model with nonideal mixing. (a) Entire studied range of composition. (b) Magnified view of the water-rich region corresponding to the shaded area in panel (a), compared with water activity derived from vapor pressure measurements (open diamonds) from ref. [58].

As for the confined DES solutions, it is striking that the evaluated values of water activity are different from those of their bulk counterparts. For pore radius $R_P = 4.15$ nm, the apparent deviation from ideal mixing is reduced. And for stronger confining condition, $R_P$ = 1.8 nm, the activation coefficient even becomes larger than unity. This observation is consistent with the reduced cryoscopic effect seen in Fig. 3 for MCM-41, with the liquidus being located at temperatures above the thermodynamic predictions based on classical Gibbs-Thomson and ideal mixing approximations.

We have recently reported similar features for confined aqueous solutions of glycerol, although this first observation seemed restricted to smallest pore size $R_P$ = 1.8 nm.[37] Understanding the origin of this phenomenon is challenging, and different interpretations can be invoked. First, it can be questioned the hypothesis classically made in the derivation of the Gibbs-Thomson thermodynamic theory. The extended version of the Gibbs-Thomson law including cryoscopic effect (eq. 3) involves the solid-liquid surface tension, which was approximated by that of bulk water. Despite the large surface curvature found in nanopores, this approximation has been commonly found to work well for pure liquids. [55] This approximation is more questionable for DES aqueous solutions, due to possible solute effects on the ice interfacial properties. However, this is unlikely to be the dominant reason for observations already made for extremely diluted solutions (i.e. < 5 % of DES molar fraction).

A second point of interest concerns the Gibbs free energy of water. We have circumvented limitations of the classical approximation made in the Gibbs-Thomson formalism (eq. 2), by using an advanced description of the water thermodynamic properties.[56] However, the extremely deep depression of the melting point found in confined DES aqueous solution implies entering into the water 'no-man's land' below 236 K, where only extrapolated values of the Gibbs free energy of water are available. As such, our results might reflect the puzzling nature of water in the low temperature range where thermodynamic anomalies are invoked, and are the subject of constant active discussion in the literature.[59]

Finally, we find it very likely that, due to pore surface interactions and spatial restrictions at the molecular scale, DES aqueous solutions adopt a different nanostructure in pores. This interpretation finds support in recent studies of the nanostructure of DESs at solid interfaces.[60] Atomic force microscopy studies of choline chloride based DESs interacting with different types of flat surfaces including Pt(111), mica and highly-ordered pyrolytic graphite, have demonstrated both lateral and vertical (layering) nanostructures.[61, 62] The interfacial liquid ordering was shown to gradually decay on the nanometer-scale as a function of the distance to the solid surface. Remarkably, the interfacial layering was also shown to be reinforced by the addition of up to 40wt% of water.[61] It is also noteworthy that, in the presence of surface-charged mesoporous silica colloids, long-range fluctuations of the structure of ethaline were measured on distance exceeding hundreds of nanometers by small angle X-ray scattering.[63]

When DESs are confined in nanoporous media, interfacial phenomena play a predominant role due to a high surface-to-volume ratio, which can greatly perturb the structure of the liquid. In fact, our finding that the chemical activity in the confined solution does not follow the bulk property of the solution at the same macroscopically averaged composition can be seen as the indication of concentration inhomogeneities resulting from the spatial segregation between water and DES constituents in pore. This phenomenon has been first discussed in a thermodynamic study of a different type of aqueous solution (glycerol) for $R_P = 1.8$ nm, but for $R_P = 4.15$ nm, bulk-like water chemical activity were recovered.[37] This is unlike the present DES systems, which already exhibit thermodynamic deviations for relatively large pore size, indicating the more complex nature of the nanostructure of DESs compared to alcohol solutions.[11, 12] Inspiring future works, it is noteworthy that, unlike confined organic binary mixtures, [49-52, 64] a direct structural examination of this phenomenon remains to be established for nanoconfined DESs.

## Conclusions

While the effects of confinement on the physical properties of conventional liquids have been the subject of intense activity over the past two decades, the topic is still in its infancy regarding emerging new classes of alternative solvents. We have performed, for the first time, a comprehensive thermodynamic study of the phase behavior of aqueous solutions of the prototypical ethaline deep eutectic solvent when confined in well-defined nanochannels.

Crystallization from neat DES and from lightly hydrated DESs can be avoided using conventional thermal conditions over an extended temperature range, both in bulk and in confined geometry. In fact, they form good glass-forming systems, and the impact of confinement on their glass transitions is moderate. This is of major practical interest, and more specifically, it means that the beneficial effect of water on the dynamics of DES is preserved in nanoscopic environment.

For hydration level beyond a threshold value $W_g' \approx 30\%$, DES aqueous solutions phase separate into ice and a maximally-freeze concentrated DES solution. The liquidus exhibits an important dependence on both the pore size and the DES composition. This illustrates the confinement and cryoscopic effects, which are classically expressed in terms of the Gibbs-Thomson and Raoult's laws, respectively. For bulk solution, and in a lesser extent for the large pore size ($R_P = 4.15$ nm), we demonstrate that predictions from classical thermodynamics are in good agreement with experimental observations, as soon as non-ideal mixing effects and an advanced description of Gibbs free energy of water are implemented. However, we conclude that such a classical thermodynamic approach fails to capture the phase behavior of DES solutions in the small pore size ($R_P$ = 1.8 nm). Understanding the origin of this discrepancy remains a very challenging question.

In fact, due to the extremely deep melting depression attained under these conditions, the liquidus moves downward into the 'no-man's land' of water, where the advent of thermodynamic anomalies is invoked, and only extrapolated values of the water properties are accessible. Bearing in mind this intrinsic difficulty, we infer that it could also indicate that the apparent activity of water in confined DESs is actually larger than in their bulk counterparts. As such, this remarkable feature indirectly points to the possible existence of specific nanostructures, such as concentration inhomogeneities and intermolecular correlations in DES when manipulated in nanochannels or at interfaces with solids. Given the obvious applied interest of the use of DESs in nanoporous materials, these results may serve as inspiration for future studies.

## Author statement

Benjamin Malfait: Investigation, Writing - Review & Editing.

Aicha Jani: Investigation, Writing - Review & Editing.

Denis Morineau: Conceptualization, Investigation, Writing – Original, Draft, Supervision, Funding acquisition.

## Declaration of competing interest

The authors declare that they have no known competing financial interests or personal relationships that could have appeared to influence the work reported in this paper.

## Acknowledgments

Support from Rennes Metropole and Europe (FEDER Fund – CPER PRINT$_2$TAN), and the ANR (Project NanoLiquids N° ANR-18-CE92-0011-01) is expressly acknowledged. The authors are grateful to the CNRS – network SolVATE (GDR 2035) for financial support and fruitful discussions.